\documentclass[prl,superscriptaddress,onecolumn,showpacs,amsmath,amssymb]{revtex4-1}

\usepackage{graphicx}
\usepackage{bm}
\usepackage[breaklinks=true,colorlinks,citecolor=blue,linkcolor=blue,urlcolor=blue]{hyperref}
\usepackage{color}
\usepackage{amsmath}
\usepackage{datetime}
\usepackage[normalem]{ulem}

\preprint{Submission target: Nano...? -- \today { }-- \currenttime}

\begin{document}

\title{Ionic liquid gating of InAs nanowire-based field effect transistors}

\author{J. Lieb}
\affiliation{Univ. Grenoble Alpes, CNRS, Grenoble INP, Institut N\'{e}el, 38000 Grenoble, France}
\author{V. Demontis}
\affiliation{NEST, Scuola Normale Superiore and Istituto Nanoscienze-CNR, Piazza S. Silvestro 12, I-56127 Pisa, Italy}
\author{D. Ercolani}
\affiliation{NEST, Scuola Normale Superiore and Istituto Nanoscienze-CNR, Piazza S. Silvestro 12, I-56127 Pisa, Italy}
\author{V. Zannier}
\affiliation{NEST, Scuola Normale Superiore and Istituto Nanoscienze-CNR, Piazza S. Silvestro 12, I-56127 Pisa, Italy}
\author{L. Sorba}
\affiliation{NEST, Scuola Normale Superiore and Istituto Nanoscienze-CNR, Piazza S. Silvestro 12, I-56127 Pisa, Italy}
\author{S. Ono}
\affiliation{Central Research Institute of Electric Power Industry,  Yokosuka, Kanagawa 240-0196, Japan}
\affiliation{Central Research Institute of Electric Power Industry, Yokosuka, Japan} 
\author{F. Beltram}
\affiliation{NEST, Scuola Normale Superiore and Istituto Nanoscienze-CNR, Piazza S. Silvestro 12, I-56127 Pisa, Italy}
\author{B. Sac\'{e}p\'{e}}
\affiliation{Univ. Grenoble Alpes, CNRS, Grenoble INP, Institut N\'{e}el, 38000 Grenoble, France}
\author{F. Rossella}
\affiliation{NEST, Scuola Normale Superiore and Istituto Nanoscienze-CNR, Piazza S. Silvestro 12, I-56127 Pisa, Italy}

\begin{abstract}
We report the operation of a field-effect transistor based on a single InAs nanowire gated by an ionic liquid. Liquid gating yields very efficient carrier modulation with a transconductance value thirty time larger than standard back gating with the SiO$_{2}$/Si$_{++}$ substrate. Thanks to this wide modulation we show the controlled evolution from semiconductor to metallic-like behavior in the nanowire. This work provides the first systematic study of ionic-liquid gating in electronic devices based on individual III-V semiconductor nanowires: we argue this architecture opens the way to a wide range of fundamental and applied studies from the phase-transitions to bioelectronics.


\end{abstract}

\maketitle
\section{1. Introduction}
{\it Iontronics} targets the control of electrical properties and functionality of electronic devices by exploiting $ionic$ motion and arrangement\cite{17Bisri,15Chun}. It represents an interdisciplinary field encompassing electrochemistry \cite{09Scrosati}, solid-state physics \cite{12Gonnelli}, energy storage \cite{17Dou}, electronics \cite{17Ma}  and biological sciences \cite{13cicoira}. A key element driving the functionality of these devices is the electric-double-layer (EDL) \cite{90review} formed at the interface between an (electronically-insulating) ionic conductor and an electronic conductor such as an organic system \cite{11Leger} or an inorganic semiconductor \cite{85Tardella}. The use of ionic liquids \cite{17ILs,17Futamura,02Green,16Chiappe} for the realization of EDL transistors (EDLTs) \cite{13EDLTs} was shown to yield very high local electric fields and efficient carrier-density modulation, and was first proposed for the modulation of charge carriers in oxide- \cite{07Misra,09Yuan,16GG,14GG} and organic-semiconductor systems \cite{08Ono}. More recently, this approach was applied to nanomaterials such as 2D systems including graphene \cite{17Paolucci,15Gonnelli}, layered transition-metal dichalcogenides \cite{18Iwasa,14Jo,12Braga}), quasi-1D systems such as nanotubes  \cite{17Iwasa,16Iwasa,14TE,14Minot,12Pacios}, and nanowires (NW) made from group IV and II-VI semiconductors \cite{17Appetecchi,15Pei,15Lee,15CdSe,14Cuniberti,Schone} or oxides \cite{16Peng,15Goldman}.
In particular, the ionic liquid EMIm-TFSI has been widely used for gating field effect devices fabricated starting from different nanostructures of diverse materials. This reflects the versatility of this liquid systems in terms of the physico-chemical compatibility with different solid state systems, as well as in terms of the possibility of operation in a relatively large electrochemical window \cite{maan13,weingarth13,saeed16}. To date, ionic liquid gating represents the most powerful approach to achieve field effect control of semiconductor nanostructures, in terms of the effective electric field \cite{prassides, ono2009,ono2008,bisri2017}.

The field-effect control of III-V semiconductor nanostructure devices represents a formidable tool both for fundamental studies and technological applications \cite{IEEE,Riel}. 
In this frame, NW-based field-effect devices are widely expected to impact significantly in different fields from energy harvesting and storage \cite{2018reviewRR,2016SciRep,NRzardo} to quantum-computation/communication technologies \cite{Majo,14NNross,QDfast}, from nanoelectronics\cite{16NLrocci} to sensing applications \cite{sens}. Ionic-liquid gating was proposed for GaAs-supported InAs quantum dots contacted with nano-gap source-drain electrodes \cite{2013Shibata}, while a similar approach  based on polymer electrolytes was used for InAs NWs \cite{17NLmicolich,15AdvFunMatMicolich,12NLliang}. To the best of our knowledge, the use of ionic liquids as a gate for field-effect devices based on individual NWs of group III-V semiconductors is still unreported. 

In this work, field-effect transistors (FETs) based on individual InAs NWs were realized and controlled via a voltage-biased ionic-liquid gate implemented by a droplet of ionic liquid surrounding the nanodevice. Electric transport measurements were carried out in a wide range of source-drain voltage bias values ($V_{DS}$), while the voltage $V_{LG}$ applied to the liquid gate was changed within few Volts around zero bias. The transconductance was also probed in a different configuration. We used the SiO$_{2}$/Si$_{++}$ substrate as a back-gate to which we applied a voltage-bias $V_{BG}$ in the range +30 to -30 V. InAs NW-based FET upon ionic-liquid gating is demonstrated, and we found a superior performance of liquid gate operation with respect to standard SiO$_{2}$/Si$_{++}$ back-gating. We measured the temperature dependence of the NW resistance from 240 K to 4 K for different liquid-gate applied voltages: semiconducting or quasi-metallic behavior was observed depending on the $V_{LG}$ applied. Moreover, we highlighted the impact of temperature and gate-voltage sweep rates on the hysteresis observed in the transconductance upon forward and reverse gate sweeps. We shall argue that the present results open the way to the exploitation of ionic-liquid gating in nanodevices based on III-V semiconductor NWs, and suggest new avenues to investigations spanning from nanoscale thermoelectrics to subwavelength light emission or detection, and from phase transitions to bioelectronics.
Besides, the unique capability of the ionic liquid environment in terms of tuning the effective electron density \cite{prassides, ono2009,ono2008,bisri2017} combined to the transport properties of III-V semiconductor nanostructure and in particular nanowires, are of undoubted relevance for fundamental studies. In particular, our results indicate that ionic liquid gating can provide a formidable tool for the advanced field-effect control of a material system - InAs nanowires - that is emerging as an ideal candidate for the experimental study of the physics of Majorana fermions in nanodevices \cite{mourik12,Lutchyn2018,Antipov2018}.

\section{2. Results and Discussion}
\subsection{2.1 Realization of field effect transistors based on single InAs nanowires gated by ionic liquid}
Devices were fabricated starting from wurtzite n-type InAs NWs grown on (111)B InAs substrates (Fig. 1a) exploiting gold-catalyzed chemical beam epitaxy~\cite{2015Gomes}  (see Supporting Information and Methods section for further details). 
NWs were mechanically detached from the growth substrate by means of sonication and dispersed in isopropyl alcohol (IPA) before being transferred to a pre-patterned $p_{++}$Si/SiO$_2$ substrate by dropcasting \cite{IEEE}  (see Supporting Information for further details).
Combs of contact electrodes (100 nm wide) were patterned using electron-beam lithography on selected InAs NWs, while square gate electrodes (100 $\mu m$ side, yellow-colored in Fig.1b) were defined in proximity of the contacted NWs. Prior to metal evaporation (Ti/Au, $10/100\,{\rm nm}$), the NW contact areas were passivated using a standard ammonium polysulfide (NH$_4$)$_2$S$_x$ solution in order to promote the formation of low-resistance ohmic contacts. Scanning electron micrographs of one of the fabricated devices are shown in Fig.1b and Fig.1c. In the latter, the InAs NW (red-colored) is contacted with a comb of four electrodes defining three NW sections that can be individually tested. Each fabricated device - contacted NW plus gate electrode - was covered with a droplet of ionic liquid, as schematically shown in Fig.1d and reported in Methods. 
Figure 1e shows a schematic cross-sectional illustration of a complete liquid-gated NW-based FET. Devices with different lengths ($L$) and widths ($W$) were realized. Typically, $L$ was chosen between 800 nm to 1200 nm, while $W$ spanned from 50 nm to 80 nm. 
We have fabricated 6 chips, each hosting 4-6 device fabricated starting from individual nanowires, for a total number of 28 devices. The SEM analysis shown evident defects in 3 devices (8 individual sections), due to misalignment in the fabrication, that avoided electrical tests. These problems were easily successively solved by optimizing the alignment procedure. The devices without defects were tested in absence of the ionic liquid drop, using a probe station at room temperature. Electric contact issues were noticed for two full devices (none of the sections were conducting) and for other 5 individual nanowire sections of different devices. We attribute these problems to the aging of the passivation solution, which was solved by reducing he utilisation period of the same solution. For all other nanowire sections, no contact issues were observed and we measured electrical resistances of the order of several kOhms.
When a positive voltage $V_{LG}$ is applied to the gate electrode immersed in the ionic liquid, the latter dresses the nanowire surface with cations and thus induces accumulation of negative charges (free electrons). Simultaneously, a source-drain voltage $V_{DS}$ (not indicated in Fig.1d) is applied between two consecutive metal electrodes defining one of the NW portions, and the resulting source-drain current $I_{DS}$ is measured. Experiments were performed in a two-wire configuration, using a multi-channel source-meter-unit to measure current in the NW channel, apply gate voltage bias and monitor leakage current. The SiO$_2$/Si substrate could be voltage-biased ($V_{BG}$) and thus implement a back gate. The back gate was grounded while operating the liquid gate. Electrical measurements were performed from room temperature down to $T = 4.2\,{\rm  K}$, in the dark and in the presence of low-pressure He exchange gas.
After bonding and deposition of the ionic liquid drop, we tested more than 5 devices from different chips, probing several nanowire sections for each device. The statistics of our main results is reported in the Supporting Information [SI-1. “Transport properties of InAs nanowire-based FETs operated via solid state back-gate”, Table 1, page 4; SI-3. “Ionic liquid gated InAs NW FETs - additional data: room temperature and low temperature device operation”, page 5-6-7, figures SI-5-6-7].

Both experimental and numeric approaches have shown that the electrochemical window of the ionic liquid used in this work ranges between about $\pm$2 V \cite{maan13,weingarth13,saeed16}. However, as the reaction rate decreases exponential with decreasing temperature, following the Arrhenius law, much higher gate voltages can be applied to the ionic liquid at temperatures slightly above the freezing point, without risking damages from electrochemical reactions. To make sure that our device stayed free from electrochemical effects, we performed cyclic voltammetry measurements through the ionic liquid at the temperatures relevant for our measurements (see Supporting Information).

\begin{figure}[h!]
\begin{center}
\includegraphics[width=0.93\textwidth]{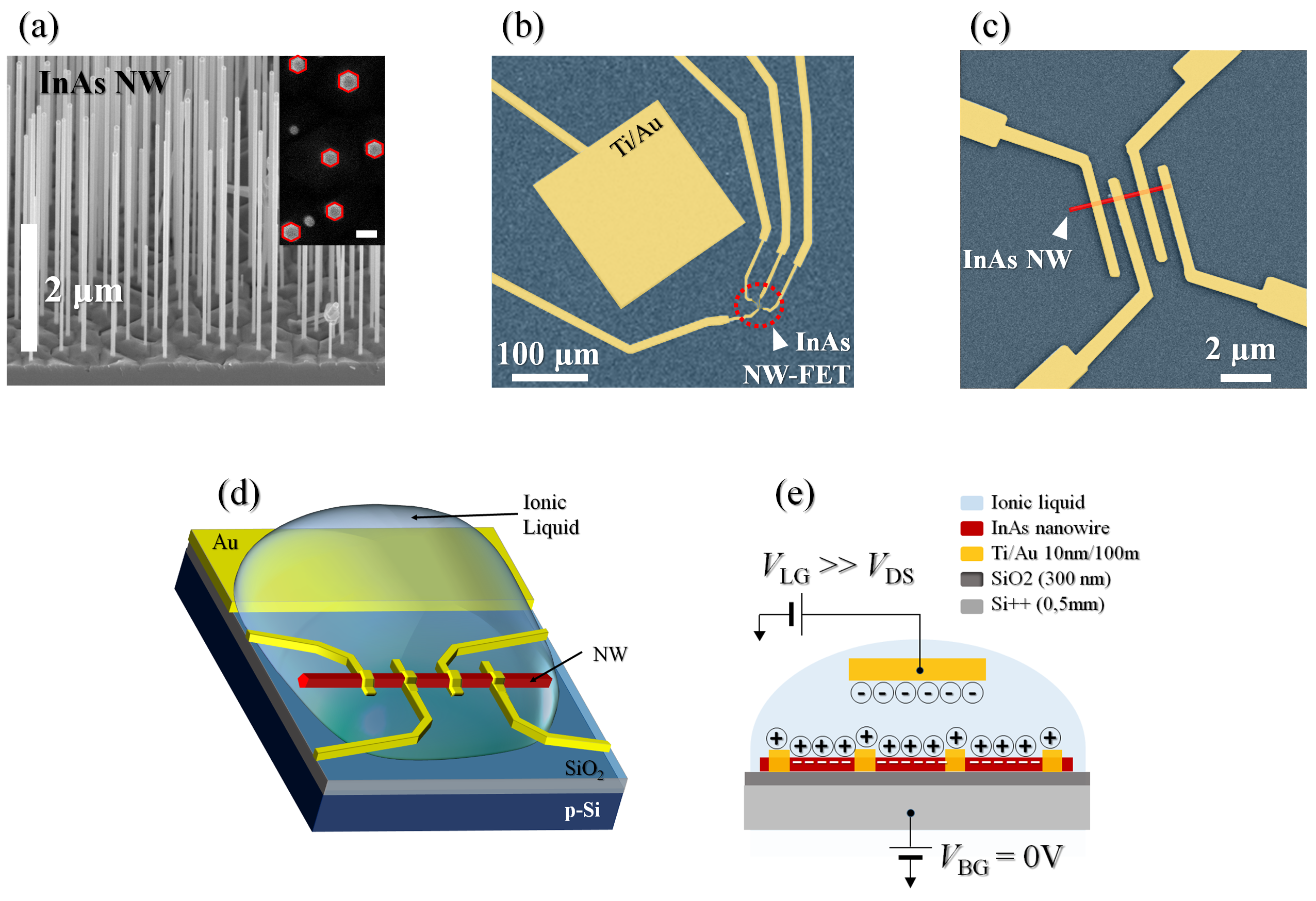}
\caption{{\bf Realization of ionic liquid-gated InAs nanowire-based FETs. } 
(a) Scanning electron microscopy (SEM) micrograph of the as-grown nanowire sample: 50$^{\circ}$C tilted and top  view (inset, scale bar is 50 nm). The NWs have an hexagonal cross section with diameters in the range 45-65 nm and average lenght around 4 um. (b) SEM micrograph (top view) of a prototypical device: the electrode used for biasing the ionic liquid is yellow-colored. The red circle indicates the position of the InAs NW-based device.
(c) Details of an InAs NW-based device: four electrodes define three NW sections. The InAs NW is red-colored.
(d) Pictorial view of an ionic liquid gated InAs NW-based FET.
(e) Schematic cross-section of a ionic liquid gated InAs NW-based FET, highlighting the operation principle.
 }
\label{fig:Cartoon}
\end{center}
\end{figure}

\subsection{2.2 Operation and performance of ionic liquid gated InAs nanowire transistors}
We report in Fig.2 the operation of a liquid-gated InAs NW-FET at 240K.
Panel (a) reports the transistor output characteristics and demonstrates successful FET operation with ohmic regime at low $V_{DS}$ bias, followed by the onset of saturation regime at high $V_{DS}$ bias. 
The low voltage bias region of the $I-V$ characteristics is magnified in panel (b) and permits to appreciate the perfectly Ohmic behaviour. From the transconductance curves reported in  panel (c), the liquid-gate threshold voltage $V_{LT}$ can be estimated. In our devices we found $ V_{T}$ ranging from  -1.8 V to -1.0 V, consistently with the relatively high concentration of n-type dopants, as discussed in the Supporting Information and later in this article.
Figure 3a reports the source-drain current $I_{DS}$ (red dots, Log scale) measured in one of our devices at 240 K as a function of $V_{LG}$ (sweep rate 4 mV/s) with applied bias $V_{DS}$ = 10 mV.
The InAs NW-based FET is driven from full saturation - passing through a region of linear response - to full pinch-off within $V_{LG}$ = +2 V and $V_{LG}$ = -2 V. 
The residual current level ($\leq 1 nA$) observed is compatible with previous results on similar ionic liquids \cite{12Braga}.
During device operation the absolute value of the leakage current measured across the gate electrode remained negligible, always below 50 pA (black dots, right $y$-scale), i.e. more than four orders of magnitude below the saturation current - and well below the off-current - for the whole range of $V_{LG}$ investigated.

The transconductance in our devices was preliminary investigated at room temperature using the back-gate without ionic-liquid droplet. Figure 3b shows the source-drain current $I_{DS}$ (blue dots, Log scale) measured at room temperature in one of the devices without ionic liquid droplet as a function of $V_{BG}$ (sweep rate $\approx$5 mV/s) with applied bias $V_{DS}$ = 4 mV. In this case leakage current did not exceed 5 nA. From the slopes $ dI_{DS}/dV_{G} $ of the linear parts of the characteristics, we can extract the electron mobility $\mu$ as 
$ \mu=dI_{DS}/dV_{BG}\cdot L^2/(C_{BG} V_{DS})$, for the back-gate configurations, with $ C_{BG} $ being the back-gate capacitance. The NW doping level was rather high ($n \approx 6\cdot10^{17}$ cm$^{3}$/Vs) and prevented full modulation of device conductance even for back-gate applied voltage intervals as large as 60 V. The highest mobility values obtained were between 135 and 600 cm$^2$/Vs (mirroring the know inter-wire non uniformity \cite{2015Gomes}). Experimental results are reported and analyzed in detail in the Supporting Information and yield estimates of the electrical mobility $\mu$ and the electron density $n$ of our NW-based devices. 

\begin{figure}[ht!]
\begin{center}
\includegraphics[width=0.99\textwidth]{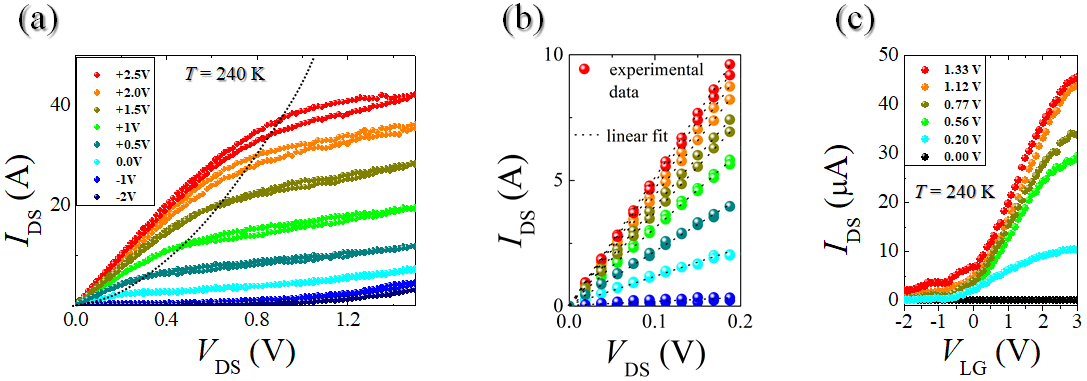}
\caption{{\bf Full operation of  ionic liquid gated InAs NW-based FET. } 
(a) Current-voltage characteristics of one of our devices measured for different liquid gate voltages. The dashed curve indicates the separation between the linear and saturation regions. 
(b) Magnification of the low bias region of datasets reported in (a), highligthing the Ohmic behavior. 
(c) Transconductance curves measured for different source-drain bias voltages.
 }
 \label{fig:Cartoon}
\end{center}
\end{figure}

\begin{figure}[h!]
\begin{center}
\includegraphics[width=1.03\textwidth]{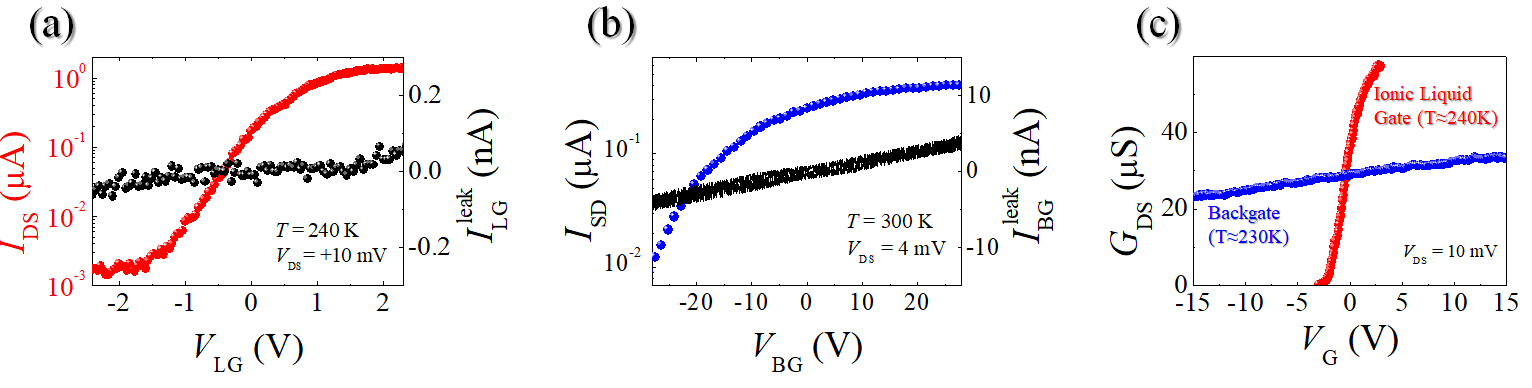}
\caption{{\bf Performance of liquid gating vs. back gating.  } 
(a) Device current (magenta curve, Log scale) and leakage current (black curve, linear scale) as a function of the applied liquid-gate voltage.
(b) Device current as function of back-gate applied voltage in absence of the ionic liquid droplet.
(c) Device transconductance measured upon liquid-gate (red circles) and back-gate (blue circles) applied voltage ($V_G$).
 }
\label{fig:Cartoon}
\end{center}
\end{figure} 

Furthermore, we can estimate the field-effect mobility using the ionic liquid gate,	provided we can estimate the gate capacitance $ C_{LG}$.
Thanks to the presence of the SiO$_2$/Si$_{++}$ substrate, our EDL-devices can be operated as "hybrid" FETs in which field effect is driven by a back-gate voltage $V_{BG}$ while the ionic-liquid landscape is kept unchanged at low temperatures.
The device operation for the two different gating configurations is reported in Fig.3c which directly compares the transconductance $G_{DS}$ of the device measured upon sweep of $V_{BG}$ (blue circles) and $V_{LG}$ (red circles). A very large difference in gating efficiency is apparent:
while a modulation of three orders of magnitude is achieved by varying $V_{LG}$ from -2 V to +2 V, the modulation of $G_{DS}$  does not exceed 15 $\%$ by varying $V_{BG}$ from -15 V to +15 V. 

Assuming $\mu$ as a parameter characteristic of the nanostructure at a certain temperature and level of extrinsic doping, and assuming  $\Delta G\propto\mu\Delta V_{G}/C_{G}$ in the linear region of the transconductance, from the comparison between the two datasets we estimated the ratio $C_{LG}$/$C_{BG}$ between the liquid-gate capacitance $C_{LG}$ and the back-gate capacitance $C_{BG}$ as $C_{LG}$/$C_{BG} \approx $ 30 in this device.
We can now estimate EDL-mobility $ \mu_{FE}=dG/dV_{LG}\cdot(L^2/C_{LG} ) $. $ C_{LG} $ is strongly material dependent and is influenced by the amount of trapped states and impurities on the device. Knowing the back-gate capacitance and the ratio of the onset slopes of the transconductance measurements done by using the back-gate and the ionic liquid gate, we can estimate the ionic liquid capacitance for each device. For back-gate operation in the sweep shown in Fig.3c we estimated $C_{BG}\approx 57aF$. This, taking into account the nanowire channel surface area $S = L\cdot 2\pi r$, with L being the length of the nanowire section and r being the radius of the nanowire, yields an ionic-liquid gate capacitance $C_{LG}\approx 0.9 \mu$F. The EDL-mobility results as $ \mu \approx 59 $ cm$^2$/V/s. All our in this way estimated mobilities range between 55 and 245 cm$ ^2 $/V/s and are therefore a factor 2 to 3 lower with respect to the estimated back-gate mobilities. This can be tentatively rationalized considering (i) our approximation of the system geometry as a cylinder onto an infinite plane, while the NW cross section is hexagonal and one of the six facets lays on the SiO$ _2 $ substrate, and (ii) possible detrimental effects on the EDL-mobility due to charge disorder induced by the ionic liquid \cite{Petach2017,Gallagher2015}. A more detailed discussion of EDL mobility and its calculation is given in the Supporting Information.

\subsection{2.3 Low-temperature behavior}
In figure 4 we report the temperature dependence of the device resistance as modulated by the liquid gate. In our study, we explored the possibility to induce a modification from semiconductor to quasi-metallic behavior in the transport properties of the nanowire-device under test. This is extremely relevant for the study of charge-density-induced phase transitions in III-V semiconductors nanostructures. 
As is shown in the following, our results provide the clear experimental evidence of a marked change in the temperature dependence of the InAs nanowire driven by the liquid-gate voltage bias, where a behavior consistent with the onset of a metal to semiconductor transitions can be observed. 
The outcome suggests that this system can be of interest to investigate quasi-1D conductivity and phase transitions at the nanoscale \cite{MIT}.
The temperature dependence of the transport properties in our devices was investigated from 240 K to 4.2 K for different gate-voltage values applied to the gate electrode. To this aim, special care was taken to allow for the stabilization of the ionic liquid and to avoid thermal hysteresis effects (see details in the Supporting Information), always checking the leakage current. Figure 4 displays the $R(T)$ curve measured in one of our devices for $T \leq$ 50 K. For $V_{LG}$ = -1.5 V (red dots) the NW is almost depleted and the $R$ $vs$ $T$ experimental curve displays a characteristic semiconductor behavior. The resistance increases abruptly in the low-temperature region (T $\le$ 30 K), with the ratio $R_{5K}/R_{200K}$ approaching the value of 100 (inset). For $V_{LG}$ = -1.0 V (orange dots), $R$ still increased quite markedly by reducing $T$ but the ration $R_{5K}/R_{200K}$ barely exceeds the value of 10.  Biasing the gate electrode with large positive voltages yielded a drastic drop in the $R_{5K}/R_{200K}$ ratio, that in fact assumes values close to one for $V_{LG}$ = +4.0 V (blue curve) and $V_{LG}$ = +7.0 V (black curve).

\begin{figure}[ht!]
\begin{center}
\includegraphics[width=0.4\textwidth]{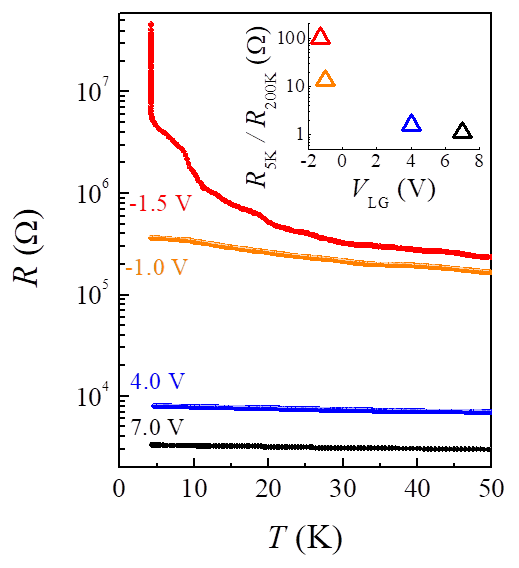}
\caption{{\bf Temperature dependence of InAs NW resistance at different ionic liquid gate voltages.  } 
 Resistance (log scale) $versus$ temperature curves measured  for different voltages $V_{LG}$ applied to the gate electrode biasing the ionic liquid, in the range from   -1.5 V (NW fully depleted) to $V_{LG}$ = +7 V (NW  fully open).
 The inset shows the ration between the resistance at 5 K  and the resistance at 200 K measured for each curve.
}
 \label{fig:Cartoon}
\end{center}
\end{figure}

\begin{figure}[h!]
\begin{center}
\includegraphics[width=0.625\textwidth]{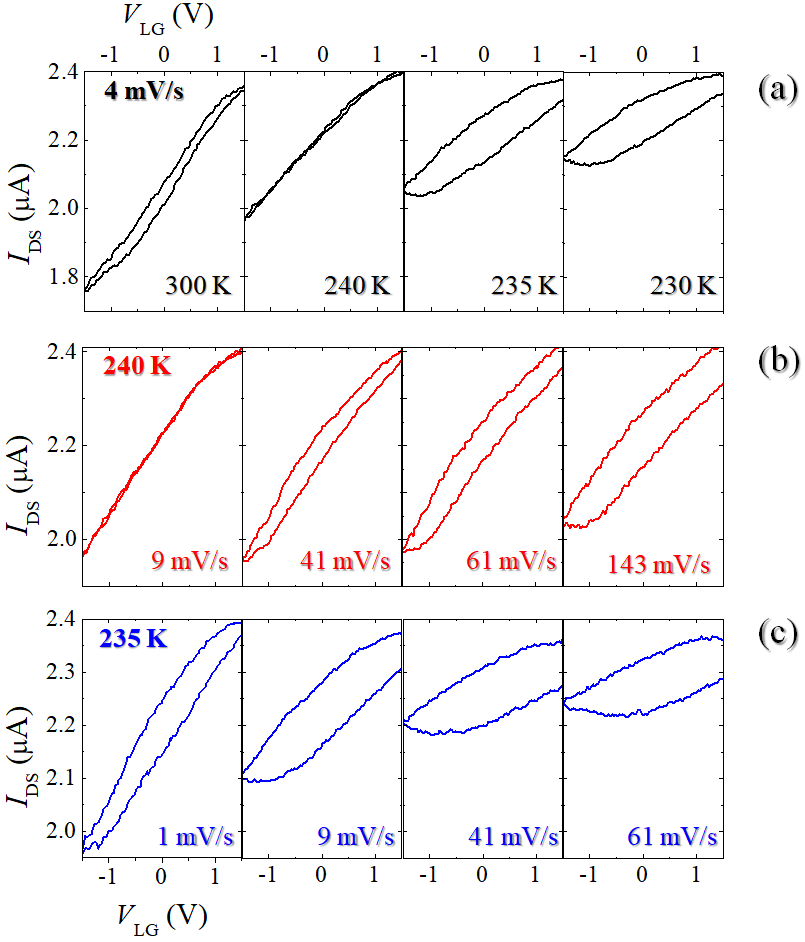}
\caption{{\bf Impact of temperature and  liquid gate voltage sweep rate.} 
(a) Forward and reverse current traces measured with gate sweep rate of 4 mV/K at different temperatures, as indicated. 
The curve at 240 K displays no hysteresis, while for temperatures above or below 240 K the forward and reverse gate sweep do not overlap. 
(b) Forward and reverse current traces measured at 240 K for different gate sweep rate, as indicated.
Hysteresis arise for sweep rate faster than 40 mV/K.
(c) Forward and reverse current traces measured at 235 K for different gate sweep rate, as indicated.
Hysteresis was present even for the slowest sweep rate.
 }
\label{fig:Cartoon}
\end{center}
\end{figure}

\subsection{2.4 Effect of temperature and sweep rate on the I-V characteristics}
Temperature and gate-voltage sweep rates can influence the measured transport properties of InAs-NW-based devices \cite{ruda}: this is particularly true for the present case of ionic-liquid gating. In general, however, a transport regime characterized by negligible hysteresis during a cycle of gate sweep is desirable: we searched for this regime in parameter space by investigating the hysteretic features displayed by our devices as a function of temperature and liquid-gate voltage-sweep rate.
Figure 5a reports current traces measured during a gate cycle ($V_{LG}$: +1.5 V $\rightarrow$ -1.5 V (forward sweep), -1.5 V $\rightarrow$ +1.5 V (reverse sweep)) for a 4 mV/s sweep rate at different temperatures from 300 K to 230 K.
Room-temperature device operation can be much affected by unwanted electrochemical phenomena leading to significant leakage current (Supporting Information) and hysteretic behavior. 
Decreasing temperature down to 240 K results in quenching the electrochemical phenomena while preserving an acceptable ion mobility. As a consequence, hysteresis decreases and virtually vanishes at about T = 240 K. Further lowering the temperature leads to degradation of ion mobility and establishment of sizable hysteresis loops. 
Figure 5b reports current traces measured at T = 240 K for sweep rates from 9 mV/s to 143 mV/s.
The increase of the sweep rate above 9 mV/s leads to a behavior change similar to that produced by temperature decrease below 240 K, i.e., hysteresis become more and more pronounced: ions can not track gate change and sign reversal for sweep rates faster than 40 mV/s.
Figure 5c reports current traces measured at T = 235 K for different sweep rates from 1 mV/s to 61 mV/s and show hysteresis with all sweep rates considered. 
This analysis shows that with the present devices optimal performance is obtained at a temperature of 240 K and a sweep rate of 4mV/s: these are the parameters used for all measurements presented in this work in Fig.s 1-4.

Finally, it is worth to stress the differences and peculiarities of the results reported in this work with respect to the outcomes of other works using a different technique to gate InAs nanowire based devices \cite{12NLliang,15AdvFunMatMicolich}. In reference \cite{12NLliang}, Liang \& Gao focus on the Rashba spin orbit coupling, widely using magnetic field for the Rashba coefficient characterization in presence of an electric field created by the polymer electrolyte, while in reference \cite{15AdvFunMatMicolich} F. Svensson and colleagues report transport properties at low temperature in the Coulomb blockade regime. Here, using ionic liquid gating, we report clear experimental evidence of the onset of a charge induced phase transition, as well as the systematic study of the gating performance and of the hysteresis. The key properties of ionic liquids exploited in this context are the negligible vapour pressure, the high thermal and electrochemical stability, the high inherent conductivities and lack of reactivity.
Finally, it has to be taken into account that the ionic liquid gating approach can be used not only for global gating but also for gating selected sections or areas of individual nanostructures. This can be easily achieve by protecting desired sections of the device with a photoresist, leaving the unprotected sections subjected to the gating action of the ionic liquid. This approach has been demonstrated in several works \cite{ueno08,ueno11,costanzo18}.

\section{3. Conclusions}
In conclusion, we investigated the device performance of a single-InAs-NW based FET with an ionic liquid as gate. To the best of our knowledge, this is the first successful demonstration of ionic-liquid gating with a NW-based device of this material of much importance for both fundamental physics and applications.
We investigated the impact of sweep rate and temperature on the stability and hysteresis of the device transconductance.
We estimated an ionic-liquid gate capacitance up to 30 times larger with respect to standard back-gate capacitance via the SiO2/Si++ substrate. This yields an increase of a factor 30 in the overall gating performance.
We studied the temperature dependence of the device resistance as modulated by the liquid gate. A clear change from semiconductor to quasi-metallic behavior was observed thus suggesting the relevance of this system for the study of charge-density-induced phase transitions in III-V semiconductors nanostructures.  
We believe the present work can stimulate novel investigations in the field of nanoscale thermoeelectrics, phase transitions, light emission from and detection by sub-wavelength semiconductor nanostructures. 
Besides, since many classes of ionic liquids are biocompatible, also bioelectronic and sensing applications of ionic liquid gated III-V semiconductor nanowires can be envisioned.
In this direction, the large surface-to-volume ratio assured by the characteristic aspect ratio of the nanowires represents a very desirable feature.

\vspace{1cm} 
\section{4. Experimental Section}

InAs NWs were grown by Au-assisted chemical beam epitaxy in a Riber Compact 21 system, employing pressure control in the metalorganic  lines to determine precursor fluxes during the growth. 
The precursors involved in the NW growth are tri-methylindium (TMIn) and tertiarybutylarsine (TBAs). 
A nominally 0.5 nm thick Au film was first deposited on (111)B InAs wafers by thermal evaporation. Before the growth was initiated, the sample was heated at $540\pm 10\,^{\circ}{\rm C}$ under TBAs flow for 20 min in order to de-wet the Au film into nanoparticles and to remove the surface oxide from the InAs substrate. 
The NWs were grown at temperature of 465$\pm$10$^{\circ}$C, with TMIn and TBAs line pressures of $\approx$0.9 Torr and $\approx$3 Torr, respectively,  for a time of $\approx$45 min.
Ditertiarybutylselenide (DtBSe) with a line pressure of 0.10 Torr was used as n-doping source.

The ionic liquid used for electrostatic doping was 1-Ethyl-3 methylimidazolium bis(trifluoromethylsulfonyl)imide (EMIm-TFSI). 
Ionic liquid droplets were placed on the devices using a 40 $ \mu $m Pt wire bent to a loop, and covered by thin ($ \approx 200\times300$ $\mu$m) glass slides to stabilize the droplet and minimize strain effects when freezing the liquid. 
The dipstick with the sample mounted was pumped at room temperature for about 2 hours before being dropped into liquid helium. 
The 4 K environment ensures cryogenic vacuum, which is essential to significantly reduce EDL-gate leakage current.
Two-terminal DC transport measurements, as well as the application of the gate voltage were performed with an Agilent B2902A source-measurement-unit.

{\bf Supporting Information.}  Transport properties of back-gated InAs NW FETs. Protocols for the measurement of the  T-dependence of device resistance. Additional data of ionic liquid gating, including room temperature device operation. Estimate of device mobility upon ionic liquid gating.

\vspace{1cm} 
\noindent {\bf Acknowledgements}

F.R. acknowledges the  support by the CNR through the RFBR France-Italy bilateral program.
B.S. acknowledges the H2020 ERC grant \textit{QUEST} No. 637815.
S.O. Acknowledges LANEF project.
L.S. acknoleges the Quantera project Supertop.

\end{document}